\documentclass[journal]{IEEEtran}

\usepackage{bm,graphicx,amsmath,cite,multirow,url,array,color,makecell}
\allowdisplaybreaks[4]
\usepackage{amsfonts}

\begin{document}
%
\title{On the Radiality Constraints \\ for Distribution System Restoration and Reconfiguration Problems}
%
%
%

\author{Ying~Wang,~\IEEEmembership{Student~Member,~IEEE,}
        Yin~Xu,~\IEEEmembership{Senior~Member,~IEEE,}
        Jiaxu~Li,~\IEEEmembership{Student~Member,~IEEE,}
        Jinghan~He,~\IEEEmembership{Fellow,~IEEE,}
       and Xiaojun~Wang,~\IEEEmembership{Member,~IEEE}
\thanks{This work was supported by the National Natural Science Foundation of China under
	Grant 51807004.}
\thanks{Y. Wang, Y. Xu, J. Li, J. He, and X. Wang are with the School of Electrical
	 and Engineering, Beijing Jiaotong University, 
	Beijing, 100044 China (e-mail:xuyin@bjtu.edu.cn)}}

\markboth{IEEE Power Engineering Letters,~Vol.~xx, No.~x, ~xxxx}%
{Shell \MakeLowercase{\textit{et al.}}: Bare Demo of IEEEtran.cls for IEEE Journals}

\maketitle

\begin{abstract}
Radiality constraints are involved in both distribution system restoration and reconfiguration problems. However, a set of widely used radiality constraints, i.e., the spanning tree (ST) constraints, has its limitations which have not been well recognized. In this letter, the limitation of the ST constraints is analyzed and an effective set of constraints, referred to as the single-commodity flow constraints, is presented. Furthermore, a combined set of constraints is proposed and case studies indicate that the combined constraints can gain computational efficiency in the reconfiguration problem. Recommendations on the use of radiality constraints are also provided.

\end{abstract}

\begin{IEEEkeywords}
distribution system restoration, feeder reconfiguration, radiality constraint, resilience.
\end{IEEEkeywords}

\IEEEpeerreviewmaketitle

\section{Introduction}

\IEEEPARstart{T}{he} distribution system is designed as a meshed network but operates radially. Maintaining radial topology is considered as a constraint for many optimization problems associated with distribution systems, such as reconfiguration\cite{mathpro} and restoration problems \cite{GM},\cite{second}. Many researchers seek to solve the optimization problems using mathematical programming methods as it can obtain the global optimum using the off-the-shelf optimization solvers. The radiality constraints should be explicitly expressed in a mathematical programming model. A set of widely used radiality constraints is the spanning tree constraint (ST) \cite{mathpro}, \cite{liyong}. However, to ensure the ST constraints resulting in radial topologies, some conditions must be satisfied\cite{marti}. To enforce a radial topology, an effective constraint set, i.e., the single-commodity flow constraints, noted as SCF0 in this letter, has been presented in \cite{planning}. In a recent work \cite{pscc}, a compact form of SCF0 is proposed for the reconfiguration problem.

In this letter, the limitation of the ST constraints is analyzed. Then a combined constraint set SCF+ST is proposed based on SCF0. The two effective radiality constraint sets are tested and compared. Suggestions on the selection of proper constraints for reconfiguration and restoration problems are provided based on the test results.


\section{Problem Formulation}
The distribution network is formulated as a connected undirected graph $\mathcal{G}=\left \langle\mathcal{N},\mathcal{E}\right \rangle $, where $ \mathcal{N}$ is the set of buses and $ \mathcal{E}$ the set of lines including tie lines. It is assumed that $\mathcal{S}$ is the set of power sources including substations and distributed generators and storages. The number of power sources is denoted by $ \left|\mathcal{S}\right|$. Assume that all lines are switchable. A radial topology should be maintained for the distribution network. A graph  $\mathcal{G_{\text{topo}}}=\left \langle\mathcal{N},\mathcal{E_{\text{topo}}}\right \rangle $ is used to represent the topology of the distribution network. $\mathcal{G_{\text{topo}}}\subseteq \mathcal{G} $ is a forest consisting of $ \left|\mathcal{R}\right|$ trees, where $\mathcal{R}$ is the set of roots. In each tree, the number of nodes equals the number of lines plus one, so $ \left|\mathcal{N}\right|=\left|\mathcal{E_{\text{topo}}}\right|+\left|\mathcal{R}\right|$.

The distribution system reconfiguration problem aims to minimize the power losses, subjecting to power flow equations, bus voltage limits, line current limits, generation/transformer capacity constraints, and radiality constraints \cite{imposing}. The decision variables include the line status 0-1 binaries and the power output of sources. 

In the distribution system restoration problem, the main objective is maximizing the number of restored loads, weighted by their priority, and the secondary objective is minimizing power loss \cite{GM}. The constraints are similar to those for reconfiguration problems. The decision variables include line status 0-1 binaries, load status 0-1 binaries, and the power output of sources. The number of sub-systems depends on the fault locations and the restoration strategy.

\section{Limitations of ST Constraints}
\subsection{ST Constraints and the Resulted Pseudo-root}
The ST constraints \cite{mathpro}, \cite{liyong} are based on parent-child relationship, i.e.,

\begin{gather}
b_{ij}+b_{ji}=a_{ij},\forall (i,j) \in \mathcal{E}
\label{equ:connect}\\
b_{ij}=0,\forall i \in \mathcal{R}, (i,j) \in \mathcal{E}
\label{equ:root}\\
\sum_{j:(i,j) \in \mathcal{E}}b_{ij}=1,\forall i \in \mathcal{N}\setminus \mathcal{R}
\label{equ:nonroot}
\end{gather}
where $a_{ij} \in \left\{0,1\right\} $ indicates whether line $(i,j)$ is connected, i.e., $ a_{ij}=1$ if line $(i,j)$ is connected, $a_{ij}=0$ otherwise; $ b_{ij}$ and $ b_{ji}$ are auxiliary variables associated with line $(i,j)$, indicating the parent-child relationship, i.e., if $j(i)$ is the parent node of $i(j)$, $ b_{ij}(b_{ji})=1$; otherwise, $b_{ij}(b_{ji})=0$.

Equation (\ref{equ:connect}) states that if line $(i,j)$ is connected, either node $i$ is the parent node of node $j$ or vise versa. Equations (\ref{equ:root}) and (\ref{equ:nonroot}) indicate that roots have no parent node while other nodes have exactly one parent node. 

\begin{figure}[!h]
	\centering
	\includegraphics[width=2.7in]{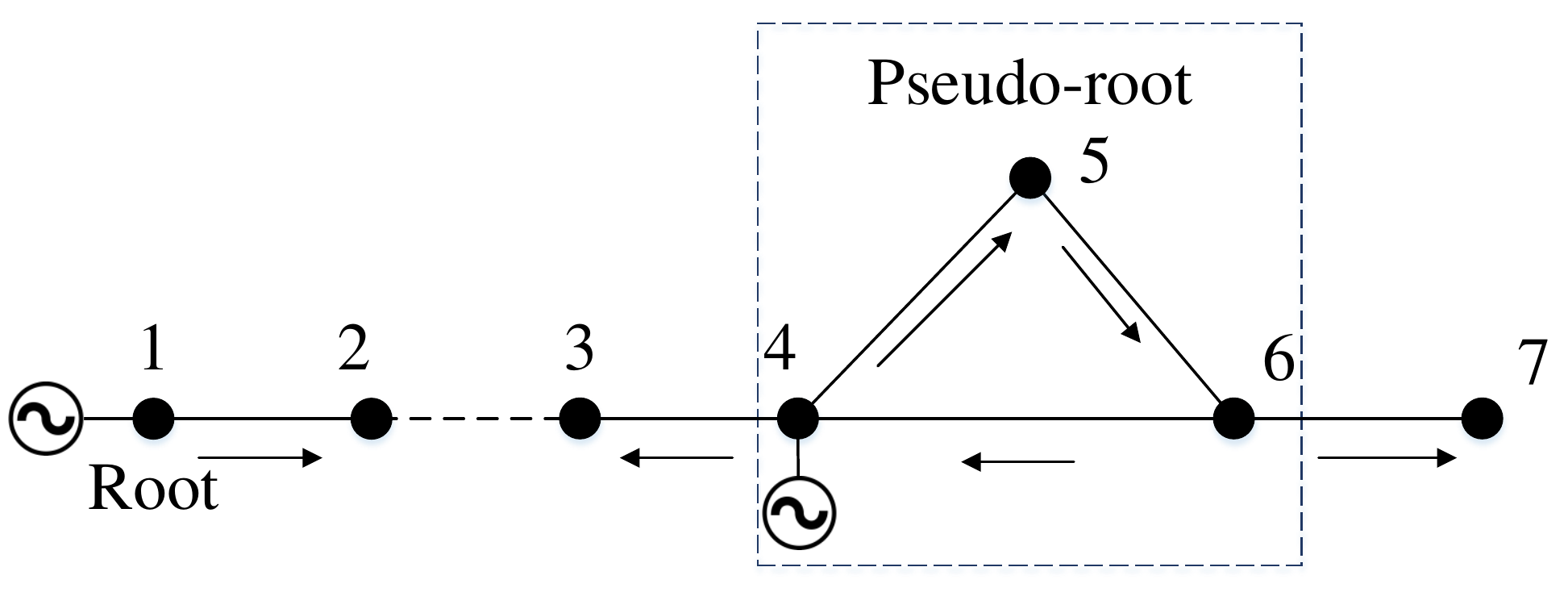}
	\caption{A post-restoration network with pseudo-root.}
	\label{fig:Pseudo-root}
\end{figure}

The ST constraints do not guarantee a radial topology when there are multiple sources (potential roots) in the distribution system. It may result in an unconnected graph containing loops, namely the \emph{pseudo-roots} , as illustrated in Fig.\ref{fig:Pseudo-root}.

The network of Fig.\ref{fig:Pseudo-root} has 2 sources and 1 loop, and 1 tree is desired with node 1 as the root. However, the network $\mathcal{G_{\text{topo}}}$ determined by constraints (\ref{equ:connect})-(\ref{equ:nonroot}) may be comprised of 2 sub-networks rather than a tree. In Fig.\ref{fig:Pseudo-root}, the arrow indicates the parent-child relationship. Node 1, which is the designated root, has no parent node and is the parent of node 2. Nodes 4, 5, and 6 form a pseudo-root, i.e., they form a loop and serve as the parent node for each other circularly. It is straightforward to validate that the ST constraints are satisfied.

\subsection{Analysis}
The pseudo-root phenomenon indicates that the set of conditions (\ref{equ:connect})-(\ref{equ:nonroot}) is not a sufficient condition for a radial topology. As stated in \cite{imposing}, the following two conditions guarantee a tree: 1) the resulted graph is connected; 2) the number of nodes is equal to the number of lines plus one. To generalize the sufficient conditions for a forest with $ \left|\mathcal{R}\right|$ trees, they can be modified as:
\\
\noindent \underline{Condition 1}: each non-root node is connected to a root directly or through other nodes;

\noindent \underline{Condition 2}: the number of nodes is equal to the number of lines plus $ \left|\mathcal{R}\right|$, i.e.,
\begin{equation}
\sum_{{(i,j)}\in \mathcal{E}}a_{ij}=\left|\mathcal{N}\right|-\left|\mathcal{R}\right|
\label{equ:node_line}
\end{equation}

The constraint set (\ref{equ:connect})--(\ref{equ:nonroot}) actually satisfies Condition 2 but not guarantees Condition 1. Therefore, the constraints (\ref{equ:connect})--(\ref{equ:nonroot}) are necessary but insufficient conditions for radial topology.

In \cite{mathpro}, power flow equations are used to guarantee Condition 1, which works if the following conditions are satisfied:
\\
\noindent \underline{Condition 3}: the number of sources is equal to that of roots, i.e., $\left|\mathcal{S}\right|=\left|\mathcal{R}\right|$;

\noindent \underline{Condition 4}: all nodes have load demand.

With these two conditions, power flow equations guarantees Condition 1 because all nodes will be connected to a root in order to serve the load demand.

As stated in \cite{imposing}, Condition 4 can be satisfied by assuming an additional small value of load (e.g., 0.001 pu) for nodes with no load, which could introduce errors in the results. In addition, condition 3 may not be satisfied in some scenarios, e.g., coordinating multiple sources to restore outage loads, where multiple sources will be connected in a sub-network.

\section{SCF-Based Radiality Constraints}

\subsection{SCF Constraints for Condition 1}
To construct SCF constraints, it is assumed that each non-root node has fictitious demand and the roots are fictitious sources of a fictitious commodity \cite{planning}, \cite{imposing}. A direction is defined for each edge in $\mathcal{E}$, making $\mathcal{G}$ a directed graph. The SCF constraints are formulated as follows:

\begin{gather}
\sum_{j:i \rightarrow j}F_{ij}+D_i=\sum_{k:k \rightarrow i}F_{ki},\forall i \in \mathcal{N}\setminus \mathcal{R}
\label{equ:fic_d}\\
\left|F_{ij}\right|\le a_{ij}M,\forall (i,j) \in \mathcal{E}
\label{equ:Fij}
\end{gather}
where $F_{ij}$ denotes the fictitious flow in line $(i,j)$; $D_i$ is fictitious demand of node $i$, which can be set as 1 for each non-root node; $M$ is a large positive number, which can be set as $ \left|\mathcal{N}\right|$.

Constraint (\ref{equ:fic_d}) requires that flow balance is satisfied for each non-root node. Constraint (\ref{equ:Fij}) forces $F_{ij}$ to zero when the line is disconnected. Constraints  (\ref{equ:fic_d})--(\ref{equ:Fij}) ensure that each node will be connected to at least one root (Condition 1).

\subsection{Two Constraints Based on SCF}
 In \cite{planning} and \cite{imposing}, constraints (\ref{equ:fic_d})--(\ref{equ:Fij}) together with (\ref{equ:node_line}) are used to guarantee the radial topology, referred to as SCF0. The constraint set proposed in \cite{pscc} is a compact form of SCF0. Equations (14a), (14b), and (15) in \cite{pscc} correspond to (\ref{equ:fic_d}), (\ref{equ:Fij}), and (\ref{equ:node_line}) in this letter, respectively.
 
Combining SCF and ST constraints can also ensure radiality, i.e., (\ref{equ:connect})--(\ref{equ:nonroot}) and (\ref{equ:fic_d})--(\ref{equ:Fij}), named as SCF+ST. The number of variables and equations needed to construct the two sets of constraints are shown in Table \ref{tab:number}.

\newcommand{\tabincell}[2]{\begin{tabular}{@{}#1@{}}#2\end{tabular}}  
\begin{table}[!htb]
	\renewcommand{\arraystretch}{1.3}
	\caption{Numbers of Variables and Constraints in SCF0 and SCF+ST}
	\label{tab:number}
	\centering
	\begin{tabular}{cccc}
		\hline		
		Constraints & Variables & Inequations & Equations   \\
		\hline
		SCF0 & $\left|\mathcal{E}\right|$  & $\left|\mathcal{E}\right|$ & $\left|\mathcal{N}\right|-\left|\mathcal{R}\right|+1$  \\
		SCF+ST & $3\left|\mathcal{E}\right|$ & $\left|\mathcal{E}\right|$ & $2(\left|\mathcal{N}\right|-\left|\mathcal{R}\right|)+\left|\mathcal{E}\right|$ \\

		\hline
	\end{tabular}
\end{table}

Compared with SCF0, SCF+ST consists of more variables and constraints. However, additional information about the relationship among line status binaries is implied in (\ref{equ:connect})--(\ref{equ:nonroot}), which may help the optimization algorithm quickly narrow down the solution search space and facilitate the solution process. Implementations for solving mixed-integer programming often relay heavily on bounds tightening to reduce feasible space \cite{minlp}. The bounds tightening process is to solve relaxed continuous sub-problems and update upper and lower objective bounds iteratively, until two bounds are identical. The relaxed sub-problem is the primal program with some of the binaries determined and others relaxed to continuous variables. For minimization problem such as reconfiguration problem, the relaxed sub-problem is solved to update the lower bound. In subproblems, the model with SCF+ST constraints may have smaller feasible region of $a_{ij}$ and obtains higher lower bound than that with SCF0, which is closer to the global optimum. 

\begin{figure}[!h]
	\centering
	\includegraphics[width=2.7in]{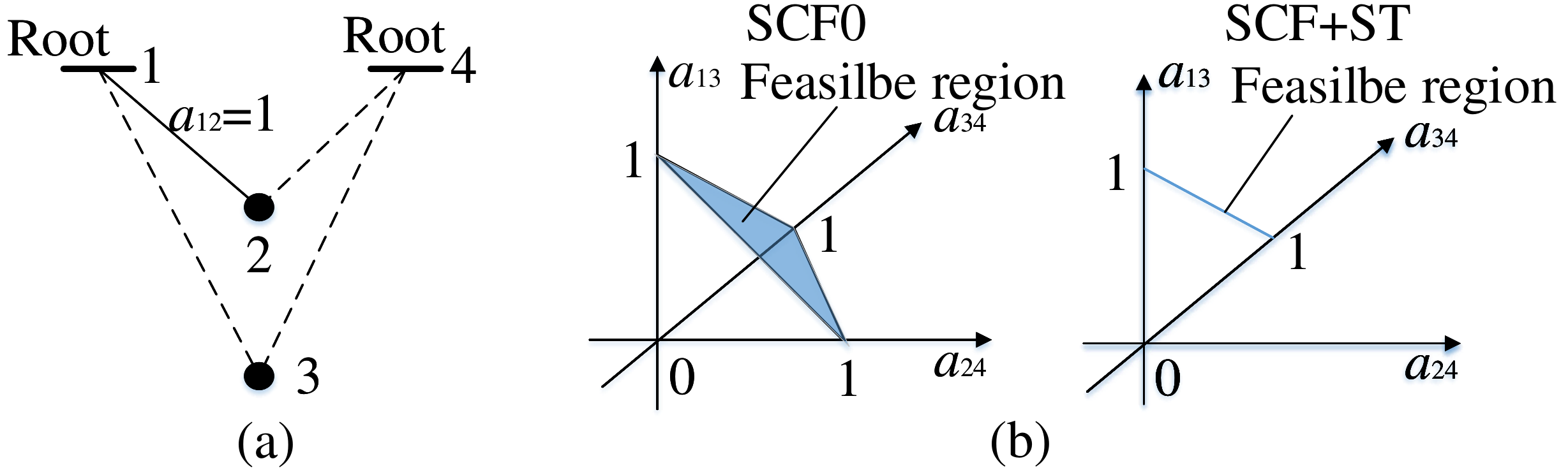}
	\caption{ Diagrams of a sub-problem and feasible region. (a) A sub-problem of a 4-node network. (b) Feasible regions of $a_{ij}$ in the sub-problem with SCF0 and SCF+ST.}
	\label{fig:demo}
\end{figure}

Take the sub-problem of a reconfiguration problem shown in Fig. \ref{fig:demo}(a) for example, where $a_{12}$ is determined as 1 and others are continuous variables. The feasible regions of $a_{ij}$ with SCF0 and SCF+ST are shown in Fig. \ref{fig:demo}(b).

The SCF0 constraints are:

\begin{equation}
\left\{
\begin{array}{lcl}
a_{13}+a_{24}+a_{34}=\left|\mathcal{N}\right|-\left|\mathcal{R}\right|-a_{12}=1 \\
0\le a_{13}\le 1\\
0\le a_{24}\le 1\\
0\le a_{34}\le 1
\end{array}
\right.
\end{equation}

While in SCF+ST, the non-root node 2 only has one parent node (constraint \eqref{equ:nonroot})  and root node 4 has no parent node (constraint \eqref{equ:root}), so $b_{24}=b_{42}=0$, then $a_{24}=0$ (constraint \eqref{equ:connect}). Therefore, the SCF+ST constraints on $a_{ij}$ are:

\begin{equation}
\left\{
\begin{array}{lcl}
a_{13}+a_{34}=1 \\
0\le a_{13}\le 1\\
0\le a_{34}\le 1\\
a_{24}=0
\end{array}
\right.
\end{equation}

The smaller feasible region can obtain higher lower bound thus narrow down the solution search space and facilitate the solution process. As a result, models with SCF+ST may be easier to solve in some scenarios.

\section{Case Studies}
\subsection{Case Information}

For the restoration problem, the modified 32-node system \cite{GM} and IEEE 123-node system \cite{coordinate1} are used for tests. There are 36 and 124 lines in the 32- and 123-node systems, respectively, indicating the numbers of line status binaries. The numbers of loads (load status binaries) are 32 and 85, respectively. In restoration problems, $\left|\mathcal{S}\right|\neq \left|\mathcal{R}\right|$, i.e., a connected sub-network may contain multiple sources. By varying locations of DGs and priority of critical loads, 300 scenarios are generated for the 32-node and IEEE 123-node systems, respectively.

For the reconfiguration problem, 4 cases, i.e., the 32-, 83-, 135-, and 201-node distribution systems \cite{REDS} are tested.

In the two problems, the power flow constraints are modeled as second-order conic constraints \cite{liyong}, resulting in mixed-integer second-order cone programming (MISOCP) models. The MISOCPs are modeled using CVXPY toolbox with Python 3. Both MOSEK solver 9.1 and CPLEX 12.8 are used to solve the models, in order to validate the consistency of test results with different solvers. Tests are carried out on a personal computer with Intel Core I7 CPU at 3.6GHz and 8GB of RAM.

The solution process will be terminated if the convergence condition is not reached after the preset maximum calculation time, because it may take unacceptable long time to find the optimal solution in some scenarios. The maximum computation time is set as 30 minutes for restoration problems (which requires online decisions) and 10 hours for reconfiguration problems (which can be solved offline). The relative optimality gap tolerance is 1e-4 for the 123-node system and 1e-8 for other systems. 
\subsection{Results}

For the restoration problems, the number of scenarios under which ST constraints fail to obtain a radial topology is 68 and 39 for the 32- and 123-node systems, respectively, while the SCF0 and SCF+ST ensure radiality in all scenarios.

\begin{figure}[!h]
		\centering
		\includegraphics[width=2.7in]{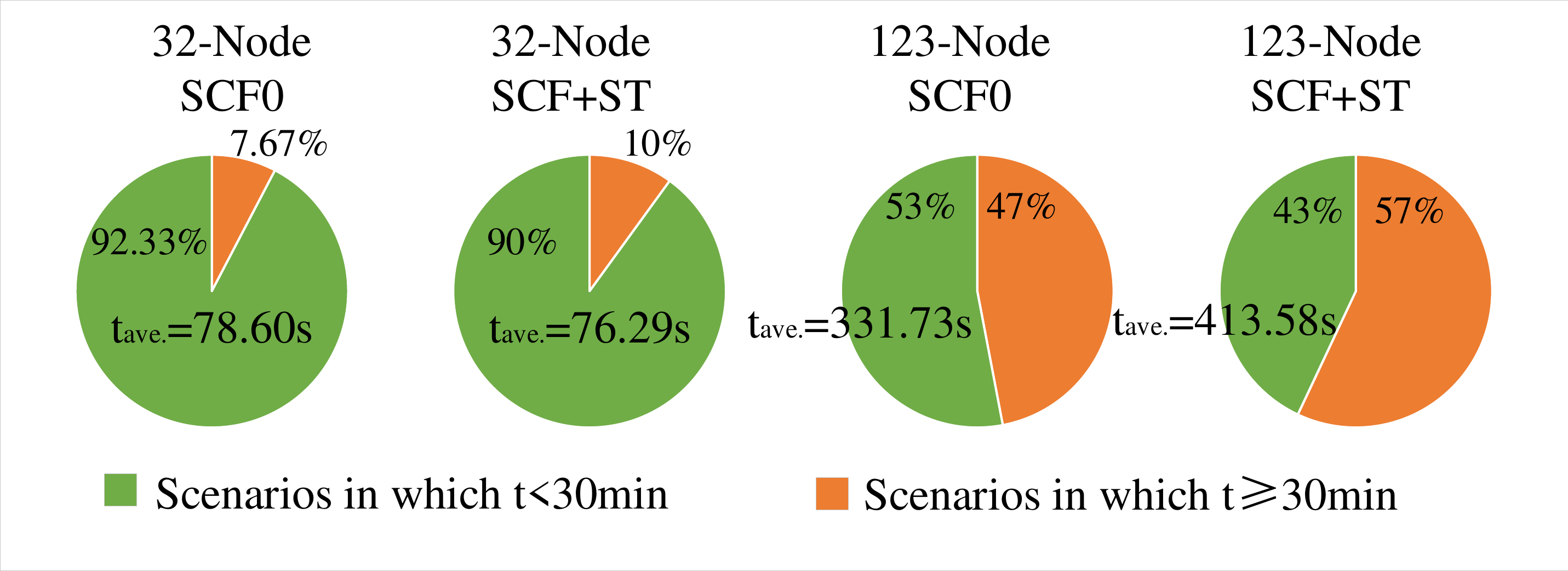}
		\caption{Test results for restoration problems solved by MOSEK.}
		\label{fig:res-result}
	\end{figure}

As to the computational efficiency of SCF0 and SCF+ST for the restoration problem, the results are shown in Fig. \ref{fig:res-result}, where $t_{\text{ave}}$ is the average computation time for scenarios that do not hit the 30-minute ceiling time.The results indicate that SCF0 outperforms SCF+ST in both average computation time and percentage violations of the ceiling time.
	
For the reconfiguration problem, the computation time is given in Table \ref{tab:Comparison}. It can be seen that SCF+ST significantly outperforms SCF0, especially for large systems.

\begin{table}[!htb]
	\renewcommand{\arraystretch}{1.3}
	\caption{Computation Time for Reconfiguration Problems}
	\label{tab:Comparison}
	\centering
	\begin{tabular}{cccccc}
		\hline
		\multicolumn{2}{c}{Case}&32&83&135&201 \\
		\hline
		\multirow{2}{*}{MOSEK}&SCF0 &4.49s&997.30s&$>$10h&$>$10h\\
		&SCF+ST &2.55s&13.55s&450.5s&6966.86s\\
		\multirow{2}{*}{CPLEX}&SCF0 &3.99s&40.38s&$>$10h&$>$10h\\
		&SCF+ST &3.77s&6.16s&438.84s&16694.8s\\
		\hline
	\end{tabular}
\end{table}

A possible reason why SCF0 and SCF+ST perform differently in reconfiguration and restoration problems is that the reconfiguration problem only contains line status binaries, but the restoration problem includes the load status binaries as well, making the relationship among binaries more complicated. Additional information contained in SCF+ST helps accelerate the solution process when line binaries are the only integer variables, i.e., in the reconfiguration problems.

\section{Conclusion}
This letter shows that the popular ST constraints are necessary but insufficient conditions for the radial topology. Both SCF0 and SCF+ST constraints guarantee radial topology, but their performance is quite different in the sense of computation time. It is recommended that SCF0 be used for service restoration problems, while SCF+ST for reconfiguration problems to achieve high computational efficiency.

\bibliographystyle{IEEEtran}
\bibliography{IEEEabrv,bibconstr}

\




\end{document}